\documentclass[]{aa}

\usepackage{amssymb}
\usepackage{color}

\usepackage{txfonts}
\usepackage{graphicx}
\usepackage{natbib}
\bibpunct{(}{)}{;}{a}{}{,}

\title{Direct observation of the energy release site in a solar flare by SDO/AIA, Hinode/EIS and RHESSI}

\author{P.J.A.~Sim\~oes \and D.R.~Graham\thanks{current address: INAF-Ossevatorio Astrofisico di Arcetri, I-50125 Firenze, Italy} \and L.~Fletcher}

\institute{SUPA, School of Physics and Astronomy, University of Glasgow, G12 8QQ, UK}

\date{Received / Accepted}

\abstract{}
{We present direct evidence for the detection of the main energy release site in a non-eruptive solar flare, SOL2013-11-09T06:38~UT. This GOES class C2.7 event was characterised by two flaring ribbons and a compact, bright coronal source located between them, which is the focus of our study.}
{We use imaging from SDO/AIA, and imaging spectroscopy from RHESSI to characterise the thermal and non-thermal emission from the coronal source, and EUV spectroscopy from the Hinode/Extreme ultraviolet Imaging Spectrometer, which scanned the coronal source during the impulsive peak, to analyse Doppler shifts in \ion{Fe}{xii} ($\log T = 6.2$) and \ion{Fe}{xxiv} ($\log T = 7.2$) emission lines, and determine the source density.}
{The coronal source exhibited an impulsive emission lightcurve in all SDO/AIA filters during the impulsive phase. RHESSI hard X-ray images indicate both thermal and non-thermal emission at the coronal source, and its plasma temperature derived from RHESSI imaging spectroscopy shows an impulsive rise, reaching a maximum at 12–-13 MK about 10 seconds prior to the hard X-ray peak. High red-shifts associated with this bright source indicate downflows of 40–-250 km s$^{-1}$ at a broad range of temperatures, interpreted as loop shrinkage and/or outflows along the magnetic field. Outflows from the coronal source towards each ribbon are also observed by SDO/AIA images at 171, 193, 211, 304 and 1600 \AA. The electron density of the source obtained from a \ion{Fe}{xiv} line pair is $10^{11.50}$ cm$^{-3}$ which is collisionally thick to electrons with energy up to 45–-65 keV, responsible for the source’s non-thermal X-ray emission.}
{Given the rich observational evidence, we conclude that the bright coronal source is the location of the main release of magnetic energy in this flare, with a geometry consistent with component reconnection between crossing, current-carrying loops. We argue that the energy that can be released via reconnection, based on observational estimates, can plausibly account for the non-thermal energetics of the flare.}
{}

\titlerunning{Direct observation of energy release site of a flare}
\authorrunning{Sim\~oes, Graham, Fletcher}

\begin{document}
\maketitle

\section{Introduction}

It is commonly accepted that the energy to power solar flares comes from the coronal magnetic field. The mechanisms for a transient release of magnetic energy, through magnetic reconnection, have been intensively investigated in both theoretical \citep[e.g.][]{Sweet:1958,Parker:1963,Petschek:1964,Syrovatskii:1966,HenouxSomov:1987,LitvinenkoSomov:1993,Longcope:1996,PriestForbes:2000,AulanierPariatDemoulin:2006} and observational \citep[e.g.][]{Demoulinvan-Driel-GesztelyiSchmieder:1993,YokoyamaAkitaMorimoto:2001,GrigisBenz:2005,SuVeronigHolman:2013,DudikJanvierAulanier:2014,van-Driel-GesztelyiBakerTorok:2014} aspects. The release of magnetic energy causes plasma heating and particle acceleration, sometimes with clear evidence of a fast re-organisation of the magnetic structure \citep{LiuLiuTorok:2012,SimoesFletcherHudson:2013,ShenZhouJi:2014}. Most of the published observational studies regarding detection of the magnetic reconnection or energy release sites are related to the {\em standard model} for eruptive flares \citep[CSHKP, ][]{Carmichael:1964,Sturrock:1966,Hirayama:1974,KoppPneuman:1976}, and include: observation of cusps above the flaring loops and inflows \citep{YokoyamaAkitaMorimoto:2001,LiuChenPetrosian:2013}, hot and fast flows above flaring arcades \citep{ImadaAokiHara:2013}, flows due to gradual pre-flare evaporation and flare eruption \citep{HarraDemoulinMandrini:2005}, and slow bi-directional flows above the loop \citep{HaraNishinoIchimoto:2006}. Other works show {\em indirect} evidence of reconnection \citep{JoshiManoharanVeronig:2007} and electron acceleration at X-points \citep{NarukageShimojoSakao:2014}. 

However, CME-less, weaker events are much more common on the Sun \citep{YashiroAkiyamaGopalswamy:2006} than large eruptive events, and might be more relevant to total power output due to solar magnetic activity, although the matter is far from settled \citep{Hudson:1991}. For these smaller events, the CSHKP standard model is less applicable, as cusp-like structures and plasmoid ejections are not observed. Recently \cite{SuVeronigHolman:2013} reported remarkable SDO/AIA observations in the non-eruptive flare SOL2011-08-17, of cool inflowing loops and newly-forming hot outflowing loops, cited as strong evidence for magnetic reconnection. 

Although the overall origin of the flare energy is known and the effects of the energy release are very well observed (heating, accelerated particles, plasma eruptions, etc) the precise location of the energy release site and its physical properties are largely unknown and difficult to detect. In this paper we present observational evidence for both direct heating and acceleration of particles in a localised coronal source, which we interpret as the main energy release site for the event SOL2013-11-09T06:38~UT, a GOES class C2.7 two-ribbon flare. 
\section{Observations and data analysis} 
The flare SOL2013-11-09T06:38~UT, a GOES class C2.7 on AR 11890, located at S11W03 near the disk centre, was observed simultaneously by several space-based instruments: {\em Reuven Ramaty High Energy Solar Spectroscopic Imager} \citep[RHESSI,][]{LinDennisHurford:2002}, {\em Atmospheric Imaging Assembly} \citep[AIA,][]{LemenTitleAkin:2012} on board of the {\em Solar Dynamics Observatory}  (SDO)  and Hinode/{\em Extreme ultraviolet Imaging Spectrometer} \citep[EIS,][]{CulhaneHarraJames:2007}. There is a gradual rise in the extreme ultraviolet (EUV), soft X-ray (SXR) and hard X-ray (HXR) emissions starting about 06:24~UT, and the non-thermal HXRs peak at 06:25:46~UT. At all wavelengths imaged the flare has two short ribbons and a compact source between them, which - as we will argue - we identify as a coronal source.  We show in the following sections that this coronal source is characterised by impulsive time evolution, sudden plasma heating reaching 12--13 MK, high electron density, and fast red-shifts 
measured by Doppler shifts of \ion{Fe}{xii} and \ion{Fe}{xxiv} emission lines.

\subsection{Characterisation of the coronal source}\label{sec:coronal}

\begin{figure*}
\resizebox{\hsize}{!}{\includegraphics[angle=0]{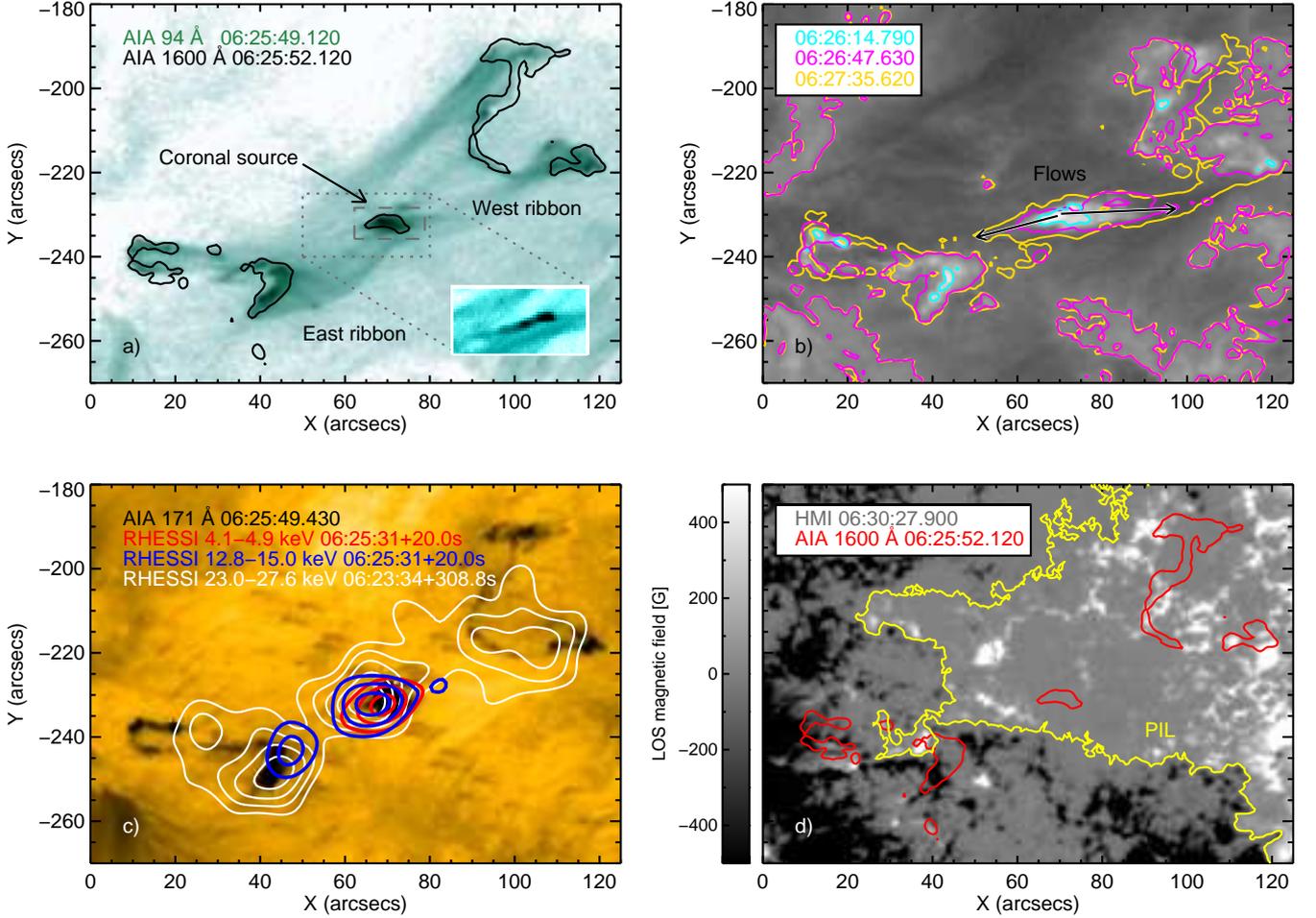}}
\caption{a) SDO/AIA 94 \AA \ image (in inverted colours) overlaid with 1600~\AA~ contours at 690 DN s$^{-1}$ pixel$^{-1}$ near the peak of the impulsive phase, showing the East and West flaring ribbons and the bright coronal source between them, indicated by the arrow. The inner frame shows the filamentary shape of the coronal source at 131~\AA~ (at 06:25:20~UT); b) SDO/AIA 211~\AA~ colour-coded { contours} time sequence of bi-directional flows away from the coronal source; c) SDO/AIA 171~\AA~ image (in inverted colours) overlaid with RHESSI images at 4.1--4.9 keV (red) and 12.8--15.0 keV (blue) near the peak of the impulsive phase { (integrated for 20 seconds)}, with contours at 50, 70 and 90~\% of the maximum of each image, { and at 23--27 keV (white, levels 50, 60, 70, 80, 90 \% of the image maximum) integrated for $\approx 300$ seconds, covering the entire main impulsive phase (time intervals indicated in the frame)}. d) SDO/HMI line-of-sight (LOS) magnetogram {(saturated at $\pm 500$ G)} 
overlaid with 1600~\AA~ contours as in a), the { yellow} line shows the magnetic polarity inversion line. Times of each image are indicaetd in each frame.}
\label{fig:aia_spot}
\end{figure*}
\begin{figure*}
\resizebox{\hsize}{!}{\includegraphics[angle=0]{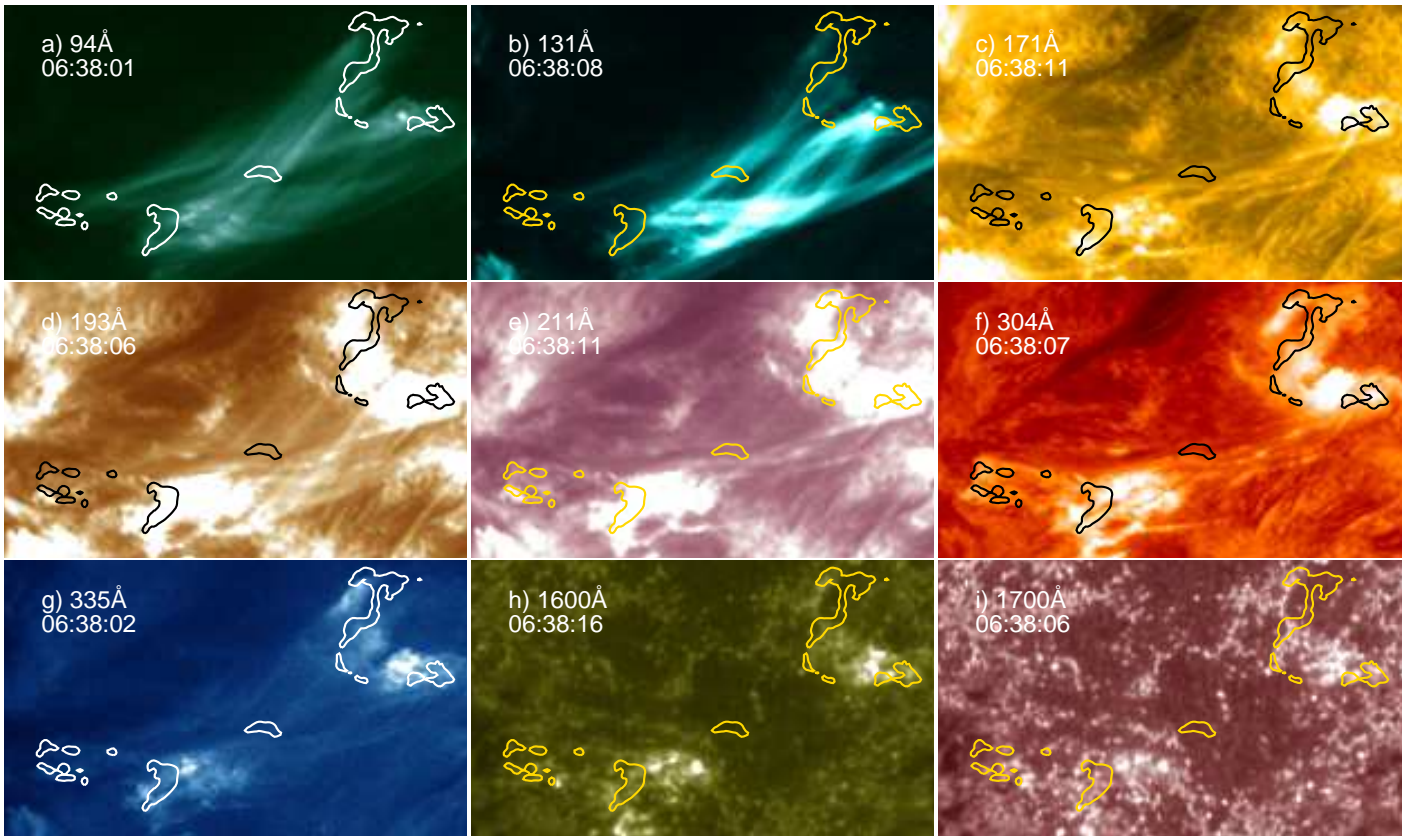}}
\caption{SDO/AIA images after the main impulsive phase, overlaid with 1600 \AA~ contours (690 DN s$^{-1}$ pixel$^{-1}$, with different colours in each panel for better contrast) at the peak of the impulsive phase 06:25:52~UT.}
\label{fig:postloops}
\end{figure*}
\begin{figure}
\resizebox{\hsize}{!}{\includegraphics[angle=0]{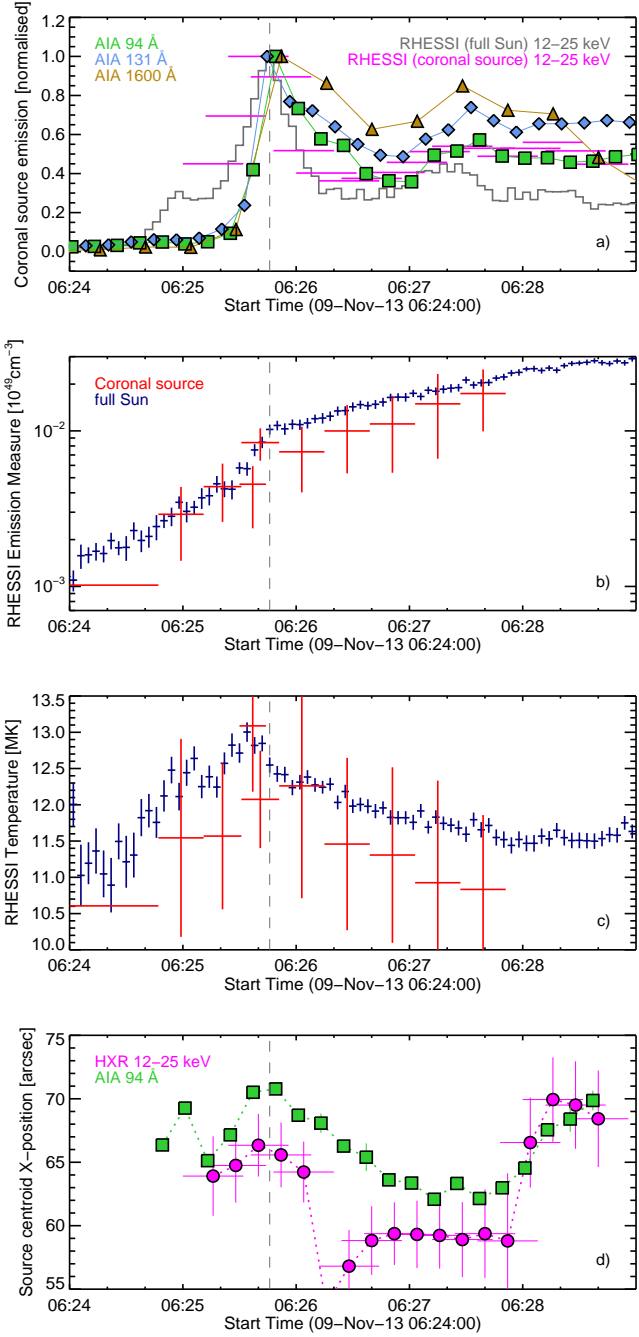}}
\caption{{\em a)} Normalised coronal source (box in Figure \ref{fig:aia_spot}a) emission at EUV 94 (green squares) and 131~\AA~ (blue diamonds), UV 1600~\AA~ (yellow triangles), HXR at 12--25 keV (magenta horizontal lines) from RHESSI images, with the { horizontal length} of the lines indicating the time interval of the image. Full Sun HXR 12-25 keV (dark gray) from RHESSI is also shown for reference. {\em b)} RHESSI emission measure (EM, cm$^{-3}$) and {\em c)} temperature ($T$, MK) from the isothermal model for spatially unresolved (full Sun) (dark blue) and from imaging spectroscopy analysis for the coronal source (red). Vertical bars show the uncertainty from the fit and horizontal bars the time integration interval. In all panels, the vertical dashed line indicates the peak time of the HXR emission above 12 keV. { {\em d)} Positions (in solar X direction) of the coronal source centroid measured from SDO/AIA 94~\AA\ and RHESSI 12--25 keV images. Horizontal bars indicate the time integration and vertical bars show a 
rough estimate of the uncertainties of the HXR centroid.}}
\label{fig:spot_emission}
\end{figure}
\begin{figure*}
\resizebox{\hsize}{!}{\includegraphics[angle=0]{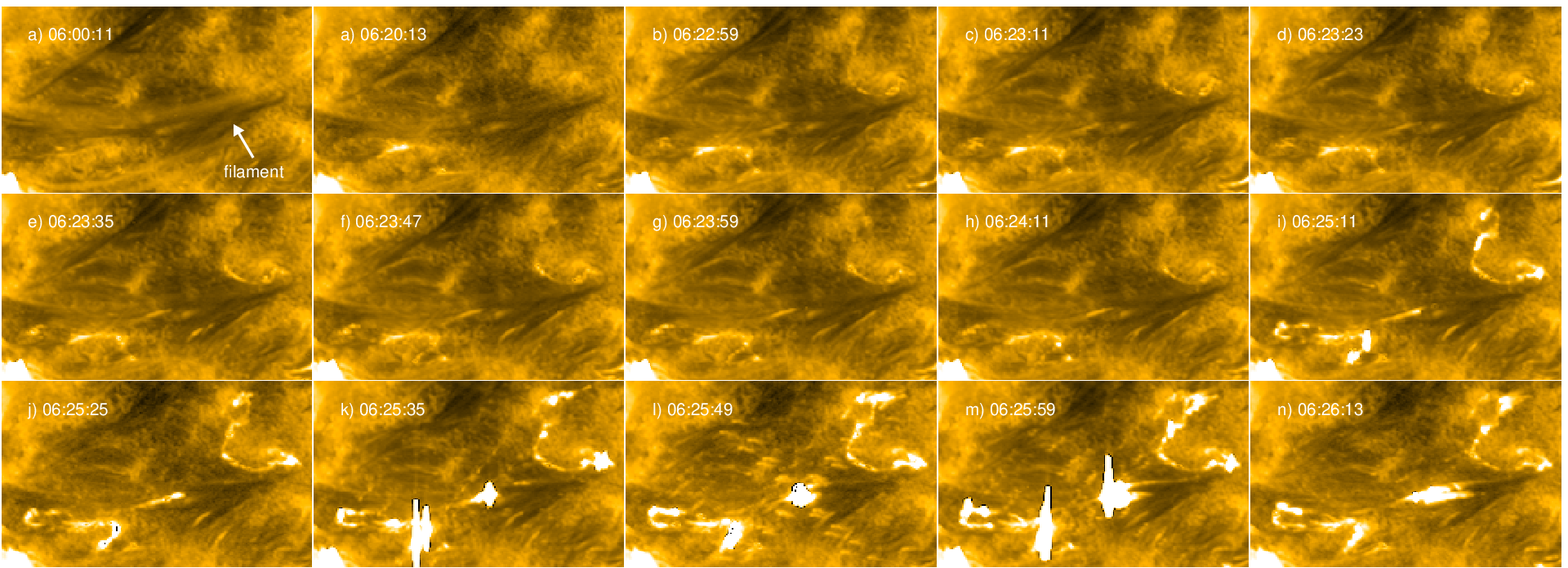}}
\caption{Time sequence of SDO/AIA 171 \AA~ images showing the pre-flare filament, the evolution of the ribbons and coronal source during the flare.}
\label{fig:filament}
\end{figure*}

The SDO/AIA provides images at filter wavelengths of 94, 131, 171, 193, 211, 304, 335, 1600 and 1700~\AA~ every 12 seconds { for the EUV filters and 24 seconds for the UV filters, with} a pixel size of 0.6 arcseconds. We use it to characterise the spatial evolution of the sources and flows in the event. The flare has two bright ribbons about 50 arcseconds apart, on opposite sides of the magnetic polarity inversion line. During the impulsive phase a remarkably bright and compact central source appears, located halfway between the ribbons: the same impulsive-phase morphology is observed in all nine EUV/UV SDO/AIA channels. Figure \ref{fig:aia_spot}a shows the SDO/AIA image at 94~\AA~ near the peak of the impulsive phase, overlaid with 1600~\AA~ contours at the level 690 DN s$^{-1}$ pixel$^{-1}$, where we identify the East and West ribbons and this central source.  Initial inspection suggested that the central source was a third footpoint, however using SDO/{\em Helioseismic and Magnetic Imager} \citep[HMI,~][]{ScherrerSchouBush:2012} magnetograms, we verified that the photospheric magnetic field directly at the source location is very weak ($\approx 13 \pm 10$~G) and featureless, looking rather like supergranular cell-centre field (Figure \ref{fig:aia_spot}d). It was not clear how a third chromospheric footpoint could be magnetically connected by the observed EUV loops to both of the other ribbons. { In fact the post-flare loops do not connect at the location of the central source, as shown by SDO/AIA images after the main impulsive phase, as shown in Fig. \ref{fig:postloops}, where hot loops connecting the ribbon regions are seen, without any connection to the region where the bright source was (as shown by the 1600 \AA~ contours at the time of the peak)}.  Moreover, its filamentary shape seen in EUV and UV, as shown 131~\AA~ at 06:25:20~UT in the panel inside Figure \ref{fig:aia_spot}a { (see also Fig. \ref{fig:filament})}, suggests a section of thin loops connecting the two ribbons. Superposed contours in a colour-coded `time sequence' of SDO/AIA 211~\AA \ images in Figure \ref{fig:aia_spot}b show bi-directional flows away from the source. These flows are also observed at SDO/AIA filters 171, 193, 304, 1600, and 1700~\AA. Later on we also show strong red-shifts in both high and low-temperature lines, much faster than are normally seen in red-shifted footpoint emission \citep[e.g.][]{MilliganDennis:2009}. We also point out the presence of dark filaments connecting the ribbon regions prior to the flare, as seen in SDO/AIA 171 ({ Fig. \ref{fig:filament}), 193, 211 and and 304} \AA, indicating that the already cool and dense coronal region is heated during the flare { \citep[e.g.][]{ChiforMasonTripathi:2006}}, explaining both the EUV/UV emission and the high density inferred from Hinode/EIS density diagnostic (Sect. \ref{sec:eis}). Taken all together, these observations lead us to identify the central source as coronal rather than chromospheric. We obtained lightcurves of the coronal source taking the sum of the pixel values inside the box shown in Figure \ref{fig:aia_spot}a, for each AIA channel. As shown in Figure \ref{fig:spot_emission}a, it had impulsive EUV/UV emission, in all AIA filters (we show 94, 131, 1600~\AA), with a stronger peak reaching a maximum at around 06:25:50~UT and a second one, peaking around 06:27:30~UT, then fading after about 06:30~UT {, as can be seen in Fig. \ref{fig:postloops}}.
\section{Evidence for the energy release site}
\subsection{A compact, moving source of heating and particle acceleration}
RHESSI imaging at non-thermal energies reveal three main sources { shown in Fig.~\ref{fig:aia_spot}c at 23--27 keV, integrated for the entire main impulsive phase ($\approx 300$ seconds)}, closely associated with both ribbons and the coronal source. At the peak of the impulsive phase however, the coronal source emission dominates at all energy ranges { as shown by the 4.1--4.9 keV (red contours) and 12.8--15.0 keV (blue) in Fig.~\ref{fig:aia_spot}c}. { At the HXR peak,} the west ribbon source is too faint to be detected by RHESSI due to its low dynamic range. The high temperature of the plasma at the coronal source suggested by the impulsive emission at 94 and 131 \AA \ is further confirmed by RHESSI imaging spectroscopic analysis. We constructed RHESSI CLEAN \citep{HurfordSchmahlSchwartz:2002} images at ten logarithmically-spaced energy bands between 3.0 and 15.0 keV throughout the impulsive phase, using front detectors 3 to 8. We employed OSPEX \citep{SchwartzCsillaghyTolbert:2002} to fit the spectra of the spatially unresolved spectra and of the coronal source source separately, giving basic imaging spectroscopy. The HXR spectra were fitted with a isothermal plus thick-target model. The plasma temperature $T$ and emission measure $\mathrm{EM}$ from the spectral fits are shown in Figure \ref{fig:spot_emission}b,c (in red) along with the values obtained from spatially unresolved spectroscopy. The most remarkable feature is the temperature that peaks at $\sim$13 MK around 06:25:40~UT (also seen in the spatially unresolved data), about 10 seconds before the HXR peak (vertical dashed-lines in Figure \ref{fig:spot_emission}).

Inspecting the HXR spectra we note that photons above 10 keV are mostly associated with non-thermal emission. A series of 12-15 keV images constructed using overlapping time intervals of 64 seconds, centred 10 seconds apart, allows us to follow the time evolution of the source positions using images with high count statistics. The centroid of the coronal HXR source and 94~\AA \ EUV source are tracked and the centroid X positions are plotted in Figure \ref{fig:spot_emission}d. The source barely moves in the Y direction, both in HXR and EUV, consistent with being situated along the loops extending in the East--West direction seen in  Figure \ref{fig:aia_spot}a. We note that the HXR and EUV centroids are not in exactly the same location, possibly due to { uncertainties in the AIA pointing}\footnote{See section 6.1 of Guide to SDO Data Analysis at \texttt{www.lmsal.com/sdodocs}}, { slight differences in aligment of different instruments (here, SDO/AIA and RHESSI)} and centroid determination at both HXR and EUV images. Nevertheless, they show clear, and very similar, displacements in the X direction: both HXR and EUV centroids shift to the East, then back to the West, spanning about $\approx 8$ arcseconds. This clear displacement of both thermal and non-thermal centroids, plus the small sizes of the EUV bright patches observed in the images (of a few arcseconds), suggests that the energy is released successively at different locations along the magnetic field, however still around a somewhat small region of around 10 arcseconds. This perhaps indicates elementary regions of energy release, and thus heating and acceleration \citep{LiuFletcher:2009,LiuHanFletcher:2010}.

\subsection{EUV Spectroscopy}\label{sec:eis}

\begin{figure}
\resizebox{\hsize}{!}{\includegraphics[angle=0]{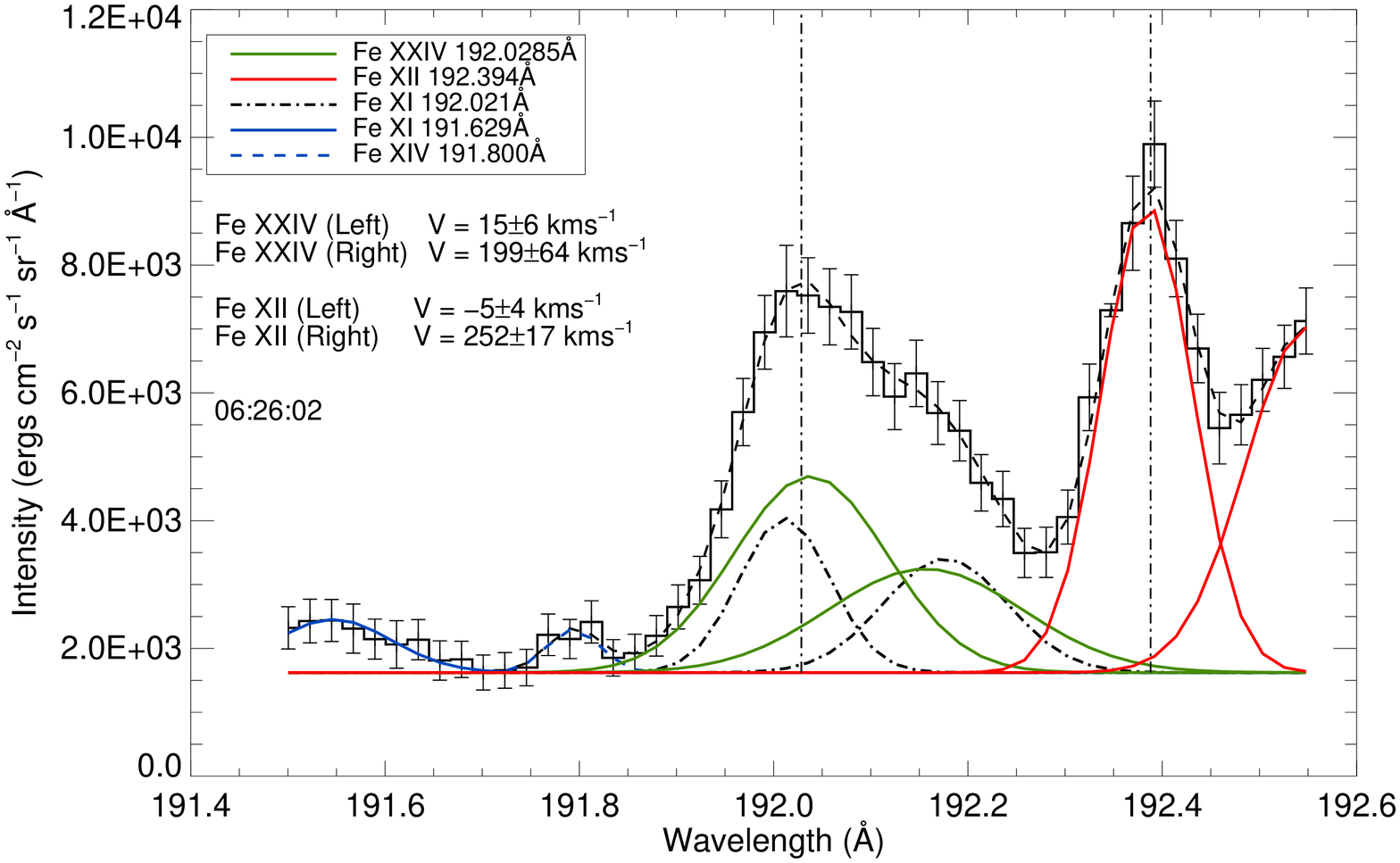}}
\resizebox{\hsize}{!}{\includegraphics[angle=0]{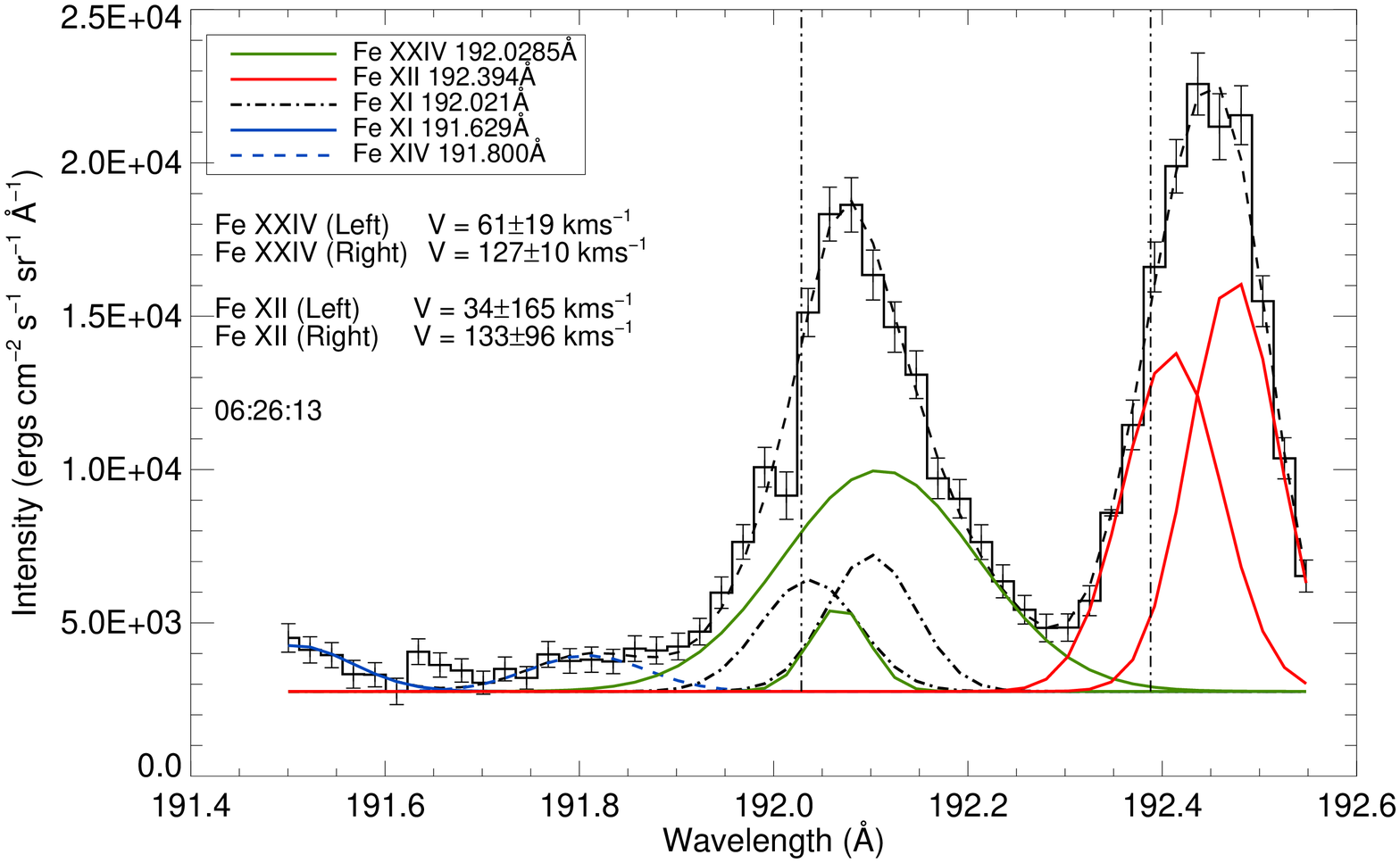}}
\resizebox{\hsize}{!}{\includegraphics[angle=0]{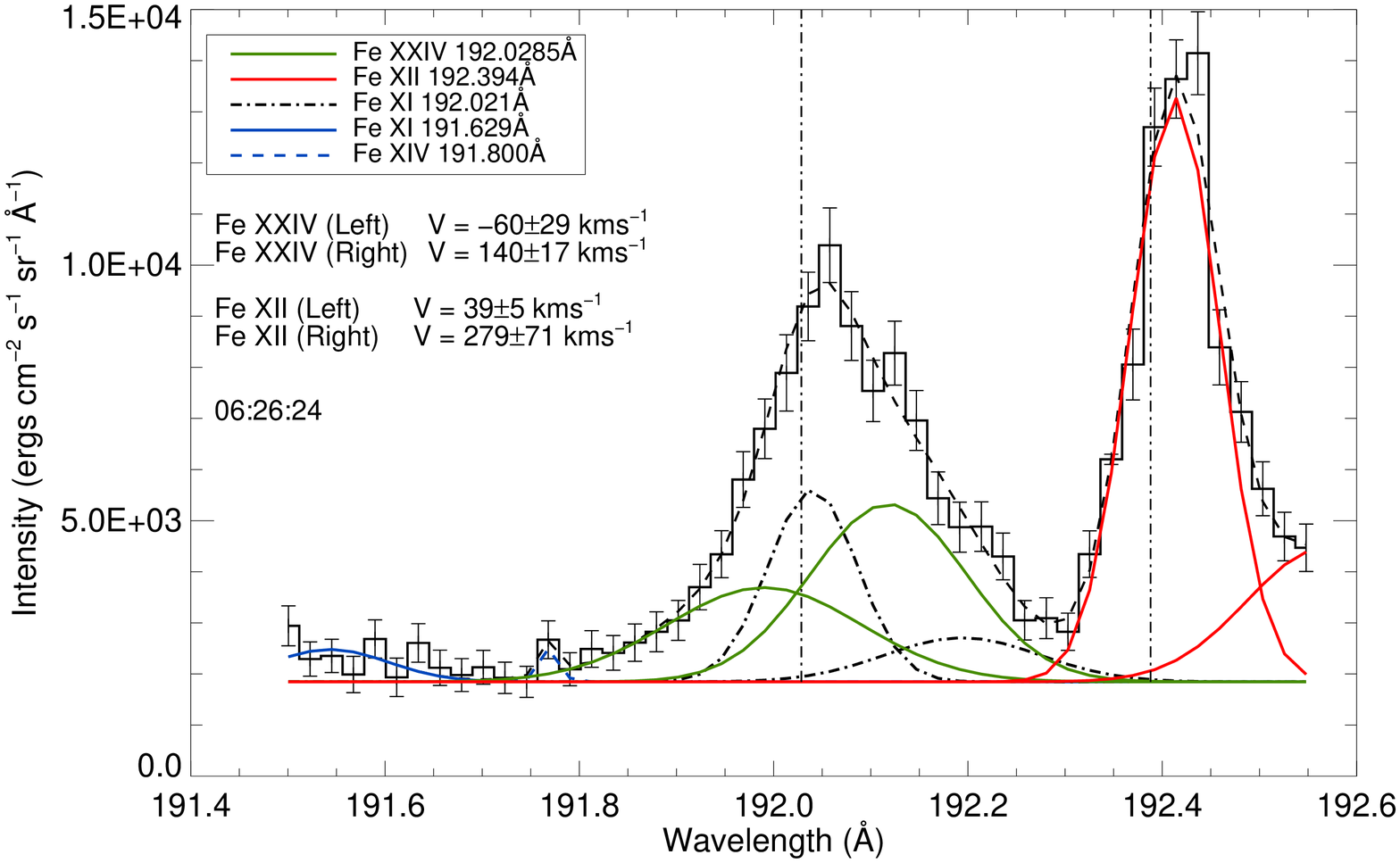}}
\caption{Fitted EIS spectra at 3 positions (and times) across the coronal source, as indicated in Figure \ref{fig:cartoon}a. The specific intensity is shown in the solid black line, with the associated data uncertainty, and the dotted black line shows the total fit to the data. Two fitted Gaussian components for \ion{Fe}{xxiv} are shown in green, for \ion{Fe}{xii} in red, and the estimated blend of \ion{Fe}{xi} in black (dash-dotted line). A vertical dash-dotted line marks the rest wavelength for the \ion{Fe}{xxiv}~192.028~\AA\ and \ion{Fe}{xii}~192.394~\AA~ lines, and Doppler velocities relative to these wavelengths for each fitted component are also displayed.}
\label{fig:eis_line}
\end{figure}

\begin{figure*}
\resizebox{\hsize}{!}{\includegraphics[angle=0]{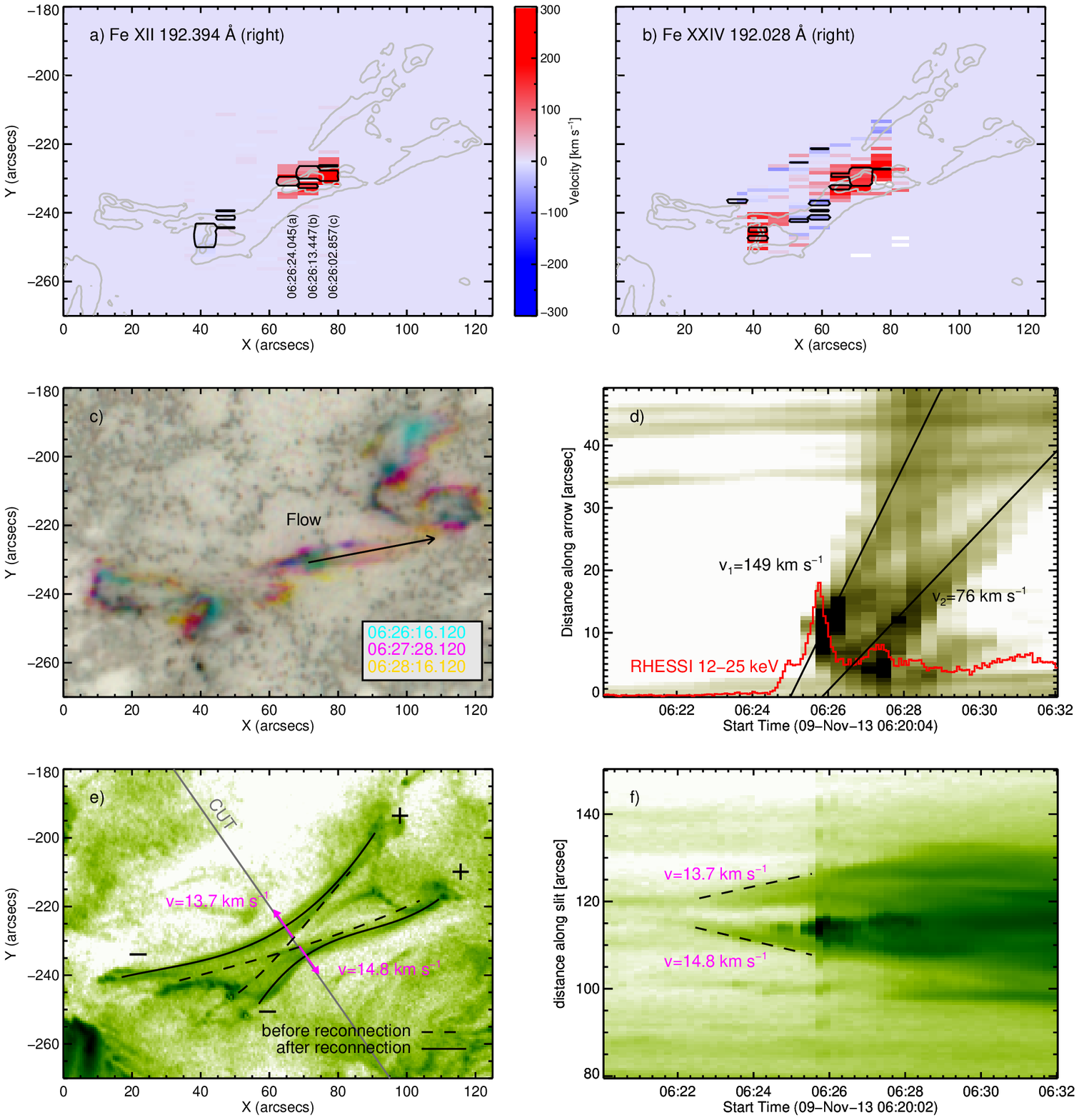}}
\caption{a) Hinode/EIS velocity maps of the red-shifed `right' component of \ion{Fe}{xii} and b) \ion{Fe}{xxiv}, overlaid with black contours at 30\% of the maximum intensity and SDO/AIA 131~\AA~ contours (in grey) for reference. c) Colour-coded time evolution of SDO/AIA 1600~\AA~ images showing plasma flows; d) Time-slice diagram along the arrow in (c), showing two outflows, originating at the coronal source, associated with RHESSI 12-25 keV peaks;  e) Representation of pre- and post-reconnection field lines; f) Time-slice diagram along the slit in (a) and estimates of the retraction speed of the post-reconnected field.}
\label{fig:cartoon}
\end{figure*}

For the entire flare duration Hinode/EIS was running the HH\_Flare+AR\_180x152 sequence, rastering the region in Figure \ref{fig:aia_spot} from West to East 9 times, each scan lasting 318 seconds. The study used the 2\arcsec\ slit with a exposure time of 9 seconds, stepping 6\arcsec\ between slit positions. The coronal source is clearly visible in the raster beginning at 06:23:34, as the spectrometer slit crossed the source at 06:26:02 and 06:26:13, 30 seconds after the non-thermal 12-25~keV HXR peak. { The EUV spectra show complex, broadened profiles with a sharp intensity rise in all observed lines, of formation temperatures between $\log T_{\rm max}=5.4$ and $\log T_{\rm max}=7.2$, in agreement with the enhancement in all AIA channels. No cooler temperature response was available in this raster set.}

A density diagnostic was carried out using a \ion{Fe}{xiv} line pair ($\log T_{\rm max}=6.3$) at 264.78~\AA~ and 274.20~\AA~ and gives $\log n_e=11.50\pm 0.82$ at the coronal source during the impulsive phase. The lower density foreground corona was removed from the diagnostic by subtracting the pre-flare spectra before calculating the diagnostic ratio. Furthermore, the density was obtained from an average of only the flare enhanced region (approx. 5\arcsec x 6\arcsec) and some pixel locations have higher densities within the region. For comparison, we found that the pre-flare density is $\log n_e=9.26\pm 0.75$. The apparent density enhancement of the region can be explained by an already dense region being heated to \ion{Fe}{xiv} temperatures during the impulsive phase. As pointed out in Sect. \ref{sec:coronal} dark, filamentary loops connecting the ribbon regions prior to the flare are observed in SDO/AIA 171 { and 304} \AA\, indicating cool and dense material in the low coronal region. { The filament and its evolution can be seen in 171 \AA, shown in Fig. \ref{fig:filament}.}

{ Significant Doppler-shifts were visible in the wide 192~\AA~spectral window which included the strong lines of \ion{Fe}{xii} ($\log T_{\rm max}$ = 6.2) at 192.394~\AA~ and \ion{Fe}{xxiv} ($\log T_{\rm max}$ = 7.2) at 192.0285~\AA. The EIS wavelength scale was calibrated using the standard SolarSoft routines to correct for drifts due to temperature changes from the orbital motion of the spacecraft, and for the slit tilt relative to the CCD. A `rest' wavelength for the \ion{Fe}{xii} line was obtained from a quiet region of a pre-flare raster which was used to measure relative Doppler velocities. Here we find a value of 192.391~\AA, which is a deviation of 3~m\AA\ blue-ward of the standard CHIANTI value. Unfortunately the \ion{Fe}{xxiv} line has no similar pre-flare signal to use as a zero-velocity reference so we use the same deviation measured for \ion{Fe}{xii}, assuming that it is systematic between lines. Nevertheless, the effect is small, on the order of $5~{\rm km~s^{-1}}$, when compared to the measured velocities.}

Figures \ref{fig:eis_line}a-c show multiple-component fitting for these lines at three slit positions across the source which are both fitted with two components; one corresponding to a stationary or slightly red-shifted component (referred to as `Left' in Figure \ref{fig:eis_line}), and a second highly red-shifted component (`Right'). A blend of \ion{Fe}{xi}~192.021 lies in the \ion{Fe}{xxiv} profile which will be significant in hot and dense regions. { The standard EIS line \ion{Fe}{xi}~188.213, nominally used to estimate such blends, was not observed in this raster.} In order to remove the \ion{Fe}{xi} contribution we first created a synthetic spectrum from the CHIANTI v7.1.3 atomic database \citep{DereLandiMason:1997,LandiYoungDere:2013} using the default CHIANTI ionisation equilibrium, a density of $10^{11}~{\rm cm^{-3}}$. We also assume a differential emission measure (DEM) obtained by \cite{GrahamHannahFletcher:2013} for flare footpoints - i.e. hot, dense and compact sources, cooling by conduction (see Section \ref{sec:thermal}) - which we expect to be very similar, at least in terms of DEM shape, to the DEM of this coronal source. Obtaining DEMs of the coronal source and ribbons will be the subject of a future study. With these assumptions the intensity ratio of \ion{Fe}{xi} 192.021 to \ion{Fe}{xii} 192.394 was predicted to be 0.31. The two \ion{Fe}{xi}~192.021 components were estimated from each of the fitted \ion{Fe}{xii} components, assuming that the widths and relative centroid shifts were equal to the \ion{Fe}{xii} parameters, and are plotted in black dash dotted lines on Figure \ref{fig:eis_line}.

Figure \ref{fig:cartoon}a,b shows Hinode/EIS velocity maps of the `Right' components for \ion{Fe}{xii} and \ion{Fe}{xxiv} in the 192 \AA\ window, revealing strong and fast red-shifts at the coronal source. The flare is located close to solar disk centre, so the measured red-shifts correspond to a downward bulk motion of 40--250 km s$^{-1}$ towards the solar surface. This speed is higher than than the downflow speeds of a few 10s of km s$^{-1}$ observed in footpoints by EIS \citep[e.g.][]{MilliganDennis:2009,WatanabeHaraSterling:2010}. The red-shifts could be caused by loop shrinkage and/or line-of-sight projection of plasma flowing along the magnetic loops towards the ribbons. The plasma flows observed by AIA (Figure \ref{fig:aia_spot}b) qualitatively reinforces the latter case. No strong red-shifts from other parts of the loops connecting the ribbons are observed, as would be expected in either of these cases, but the raster slit may simply have missed them. By verifying the position of EIS slit over AIA images, we see that the slit moves to the east before the onset of the westward plasma outflows.

\begin{figure*}
\resizebox{\hsize}{!}{\includegraphics[angle=0]{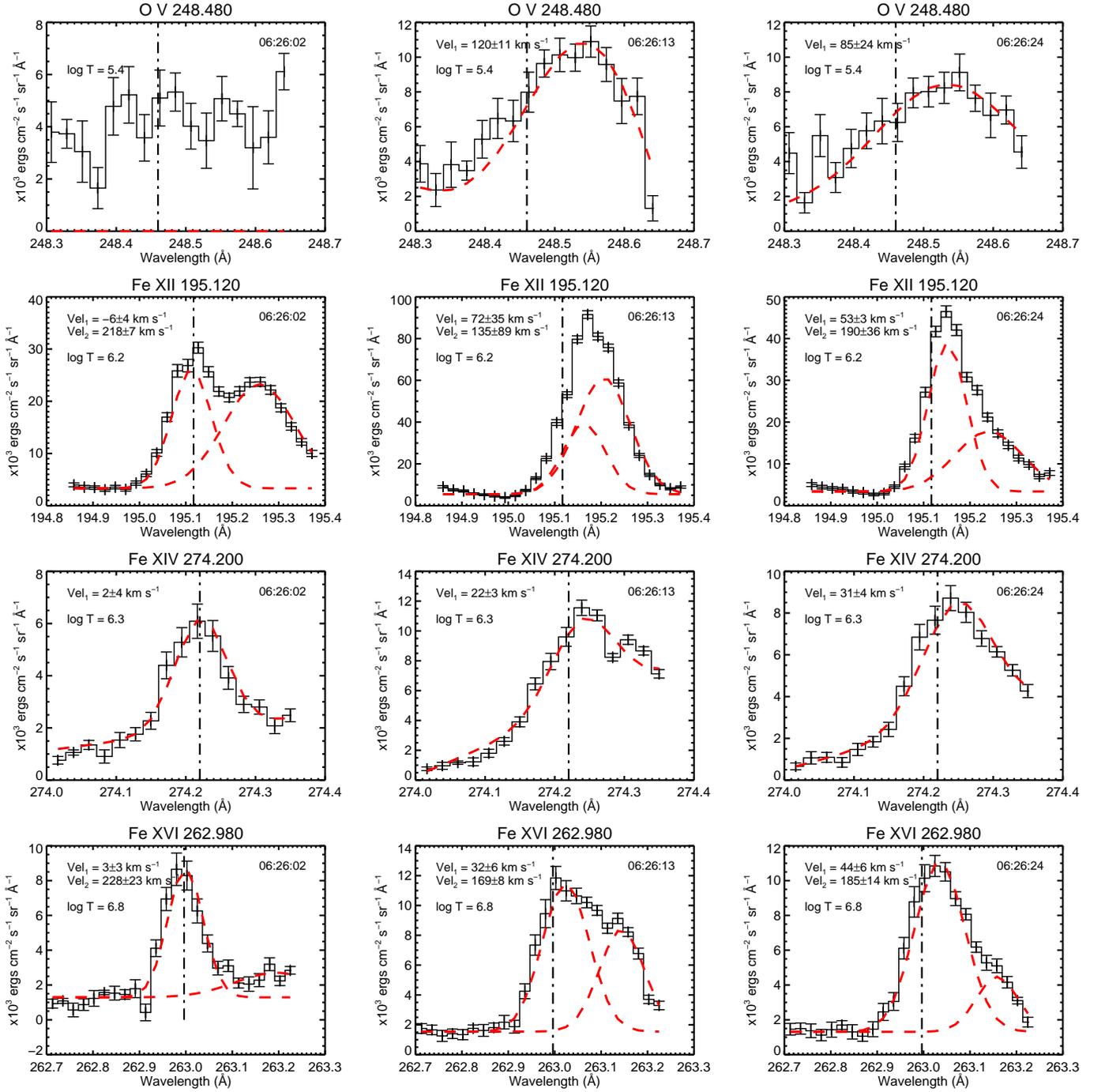}}
\caption{Fitted EIS spectra across the coronal source, as for the same positions in Figures \ref{fig:eis_line}a-c, at various temperatures. The vertical dashed-dotted line marks the rest wavelength determined from the quiet background. Rest wavelength of each ion, formation temperature and velocity of the red-shifted components are indicated in each panel.}
\label{fig:redshifts}
\end{figure*}

{ Supporting these observations of fast downflows are several other highly red-shifted emission lines with formation temperatures ranging between $\log T = 5.4 - 6.8$. Spectral profiles are shown in Figure \ref{fig:redshifts} for the same 3 slit position as in Figures \ref{fig:eis_line}. Immediately apparent are the red-shifts observed in the \ion{Fe}{xii} 195.120 \AA\ line (second row). Downflow velocities are again evident between 100-220 km/s and the relative intensities of the two Gaussian components correspond with those in the \ion{Fe}{xii} 192.394 \AA\ line. The cooler \ion{O}{v} 248.480 \AA\ transition region line at $\log T = 5.4$ also displays a red-shifted profile. The low instrumental effective area at this wavelength make fitting uncertain, however the profile is clearly shifted red-ward of the rest wavelength. Likewise, the \ion{Fe}{xiv} line (third row) has dominant emission in its red wing. In \ion{Fe}{xvi} 262.980 \AA\ a second component is distinct and can be reliably fitted, again showing red-shifts on the order of 160-180~km/s. Nowhere in the coronal source do we see any evidence for blue-shifted emission as would be expected from an evaporating footpoint \citep[see e.g.][]{MilliganDennis:2009}.}

\section{Discussion}

\subsection{Thermal properties of the coronal source}\label{sec:thermal}
The coronal source is impulsively heated, reaching 12-13 MK (from HXR data), up to 16 MK (\ion{Fe}{xxiv} emission line) and cools fast. The temperature peak is reached about 10 seconds before the HXR peak which indicates that the plasma cooling rate exceeds the heating rate before the peak particle acceleration. We can estimate the cooling time $\tau$ of the source as the thermal energy $E_\mathrm{th}$ divided by the conductive ($L_c$) and radiative ($L_r$) loss rates. RHESSI imaging spectroscopic analysis of the coronal source at the temperature peak (06:25:41~UT) gives $\mathrm{EM}=4.5 \pm 1.4 \times 10^{46} \rm{cm}^{-3}$ and $T=13\pm 1$ MK. The thermal energy content $E_\mathrm{th}=3k_bT(\mathrm{EM}\times V)^{1/2}=1.6 \sim 3.5 \times 10^{29}~\mathrm{erg}$, and the volume of the coronal source is estimated as $V=L\pi(D/2)^2$, where the length $L\approx 9\sim 18 $ arcsec and cross-section diameter $D \approx 6$ arcseconds were measured from AIA 94 \AA \ images. The conductive energy loss rate is
\begin{equation}
L_c = \kappa_0 T^{5/2} \frac{\partial T}{\partial s}A=\kappa_0 T^{5/2} \frac{\Delta T}{\Delta s}A,
\end{equation}
where $\kappa_0=10^{-6} \mathrm{erg\ s}^{-1} \mathrm{cm}^{-1} \mathrm{K}^{-7/2}$ is the Spitzer conductive coefficient, $\Delta s = L/2$, is the half the length of the coronal source, $A=\pi(D/2)^2$ is the cross-section area of the source, and $\Delta T=3$ MK, assuming that the loop temperature outside the coronal source is 10 MK. The radiative loss rate can be estimated by \citep{RosnerTuckerVaiana:1978}
\begin{equation}
L_r = 10^{-17.73} \mathrm{EM} \ T^{-2/3}, \quad (10^{6.3}<T<10^{7}K).
\label{eq:lr}
\end{equation}
We found that $L_c\gg L_r$, thus the cooling time is $\tau=E_\mathrm{th}/(L_c+L_r) = E_\mathrm{th}/L_c \approx 14\sim 38$ s, consistent with the time between the temperature peak and the HXR peak. Interpolating the CHIANTI 7.1 table for the radiative losses (for a density of $\log n = 10$ and coronal abundances) for the given EM and $T$ gives similar results as Eq. \ref{eq:lr}. The heating rate in this phase must be less than the total cooling rate of $L_r+L_c \approx 4\sim 8 \times 10^{26}$ erg s$^{-1}$, and the energy input (heating) rate to the thermal plasma must occur before the maximum of particle acceleration.

\subsection{Non-thermal properties of the coronal source}
The coronal HXR source is compact and intense, dominating the footpoint emission (Figure \ref{fig:aia_spot}c). An obvious interpretation is of a localised acceleration site, and a high local density, so that a large fraction of the electrons are completely stopped in the region \citep{WheatlandMelrose:1995}. { The electron density measured by EIS is $\log n_e \approx 11.50$ cm$^{-3}$}, and $L$ is  9--18 arcseconds. This gives a { column depth in the coronal source of $2.1 - 4.1 \times 10^{20}\rm{cm}^{-2}$}. From the relationship between stopping column depth $N_\mathrm{stop}=n_eL$ and energy $E_0$ of an electron that is collisionally stopped over column $N$ \citep{Emslie:1978,Tandberg-HanssenEmslie:2009}
\begin{equation}
N_\mathrm{stop}=\frac{\mu}{6\pi e^4 \ln \Lambda}E_0^2 \approx 10^{17}\mu(E_0[keV])^2,
\end{equation}
where $\mu $ is the electron pitch-angle cosine, $e$ is the electron charge, $\ln \Lambda \approx 20$ is the Coulomb logarithm. We then obtain the energy of an electron that is completely stopped in the coronal source:
\begin{equation}
E_0[keV] \approx \left(\frac{n_eL}{\mu_0 10^{17}}\right)^{\frac{1}{2}}
\end{equation}
For $\mu=1$ (a beam) the minimum kinetic energy that { electrons need to leave the region is $E_0=45\sim 65$ keV}. If $\mu < 1$ higher energy electrons will be stopped in the corona - for example, for a semi-isotropic distribution, the average pitch-angle cosine is 0.5, { and  $E_0=65\sim 90$ keV}. Electrons with energies less than $E_0$ will be stopped completely within the coronal source, heating the plasma. Particle trapping due to magnetic field convergence between corona and chromosphere will further enhance the coronal source intensity, as will any non-collisional pitch-angle scattering in the region, that enhances the trapping of electrons inside the coronal loops \citep[e.~g.][]{SimoesKontar:2013,KontarBianEmslie:2014}. But overall the picture is consistent with the coronal HXR source being caused by a brief burst of non-thermal particle acceleration in a dense coronal accelerator, with the electrons then stopping predominantly within that region.
 
\subsection{Flows in the coronal source} 
Fast plasma flows detected by spectroscopic observations have been reported as direct evidence of reconnection flows \citep{WangSuiQiu:2007,HaraWatanabeHarra:2011}. Recently, \cite{Brosius:2012} reported observations of a sudden rise of the emission of the Fe XIX line, formed at 8 MK, without enhancements of line emission from ions formed at lower temperatures. In addition, significant non-thermal broadening of the \ion{Fe}{xix} line was attributed to reconnection outflow or turbulence, and led \cite{Brosius:2012} to the conclusion of direct coronal heating. A pressure gradient created by heating may additionally drive a flow. In the event studied here, flows are detected by EIS spectroscopy and by AIA imaging in the 171, 193, 211 (shown in Figure \ref{fig:aia_spot}b) and 1600 \AA\ filters. Using 1600 \AA\ images, we estimated the flow speed from the coronal source towards the west ribbon by verifying the brightness evolution along the arrow in Figure \ref{fig:cartoon}c. The resulting time-position diagram is shown in Figure \ref{fig:cartoon}d where two flows can be identified with their estimated speeds: $v_1=$149 km~s$^{-1}$ and $v_2=$76 km~s$^{-1}$. The onset of both flows seem to be well associated with the HXR peaks (Figure \ref{fig:cartoon}f). Before the AIA outflows are observed, when the EIS slit first catches the coronal source at 06:26:02, the `left' components of \ion{Fe}{xii} and \ion{Fe}{xxiv} are at rest ($-5\pm4$ and $15\pm 6$ km~s$^{-1}$) while the `right' components have $v=200\sim 250$ km~s$^{-1}$ (see Figure \ref{fig:eis_line}a). At the subsequent slit positions, still over the coronal source, at 06:26:13 and 06:26:24, both Fe lines are stronger by a factor of $\approx 2$ and show `left' red-shift velocities of $v \approx 40$ km s$^{-1}$, at the time when of the AIA flow is first visible. For \ion{Fe}{xii}, the `right' component gets weaker and slower (see Figure \ref{fig:eis_line}b,c).  Assuming that the plasma velocity is composed of the EIS `left' line-of-sight velocity, $v_\mathrm{EIS}=40$ km s$^{-1}$, and AIA plane-projected velocity $v_\mathrm{AIA}=149$ km s$^{-1}$,  a simple triangular geometry yields $v_\mathrm{flow}=(v_\mathrm{EIS}^2 + v_\mathrm{AIA}^2)^{1/2} \approx 155$ km~s$^{-1}$.

\subsection{Interpretation of flows as pressure-driven} 
The high pressure in the coronal source may drive plasma outflows along the field. Assuming that the observed flows are field-aligned, the mean angle between the magnetic field at the coronal source and the line-of-sight would be $\theta=\mathrm{atan}(v_\mathrm{AIA}/v_\mathrm{EIS}) \approx 75^\circ$. Such pressure-driven flows would have speed approximately equal to the ion sound speed:
\begin{equation}
c_s=\left(\frac{\gamma Z kT}{m_i} \right)^{1/2},
\end{equation}
where for ionised hydrogen $Z=1$, $\gamma=5/3$, and $m_i$ is the proton mass. Taking the temperature of formation of the \ion{Fe}{xii} line as $T=1.3$ MK, we find $c_s=132$ km~s$^{-1}$ and for \ion{Fe}{xxiv} (formed at $T=15.8$ MK), $c_s=467$ km s$^{-1}$. Within the uncertainties, $v_\mathrm{flow} \approx c_s$ for $T=\mathrm{1.3~MK}$ indicating that the observed AIA flows and EIS (left) red-shifts could be driven by the pressure gradient between the hot, compact coronal source, and its immediate surroundings. 

\subsection{Interpretation in terms of component magnetic reconnection}
The loop geometry observed in this event cannot be explained by the CHSKP model, but may fit better with the scenario investigated by  \cite{Melrose:1997,Melrose:2004} of reconnection between crossing, current-carrying loops, in which current and magnetic flux are exchanged between loops. This is usually described in terms of quadrupolar reconnection \citep{HardyMelroseHudson:1998,AschwandenKosugiHanaoka:1999}. The black lines sketched in Fig. \ref{fig:cartoon}e represent pre- and post-flare field lines in this configuration, which are consistent with the magnetic polarities of the footpoints, and also the role of the coronal source as a possible location of local plasma energisation. Since they cross at an angle, the reconnection is between the anti-parallel components of the field (we ignore for now any internal twist in the reconnecting pre-flare loops), with an outflow speed reduced compared to the Alfv\'en speed  $v_{\rm A}$ \citep{Chae:1999}, i.e. $v_{\rm outflow} = v_{\rm A}\sin(\theta/2) = B\sin(\theta/2)/(4\pi m_p n_e)^{1/2}$ where  $\theta$ is the angle between the crossing field. \cite{Chae:1999} also gives an expression for the maximum energy, $\Delta E$ that can be released in the process, as the loops shorten:
\begin{equation}
\Delta E = 2 \biggl(1-\cos \frac{\theta}{2} \biggr) L A \frac{B^2}{8\pi},
\end{equation}
for pre-reconnection loop length $L$ and cross-sectional area $A$. The value of $B$ can be deduced from the plasma flow. A 2D planar interpretation of the geometry in Figure \ref{fig:cartoon}e would mean that $v_{\rm outflow}$ would be oriented along the cut direction shown. Time-slices of intensity along the cut in Fig. \ref{fig:cartoon}e are plotted in Figure \ref{fig:cartoon}f, and the spreading of the two ridges corresponds to a speed of $\sim 14\rm{km~s}^{-1}$. However, with the { electron number density of $\sim 3.2\times 10^{11}\rm{cm}^{-3}$ from the EIS diagnostics}, and an angle 30$^\circ$ between the dashed lines in the 2D geometry, { we obtain $B \sim 14$~G}, which is a rather small value. This implies that the geometry is not 2D. Instead we take the measured EIS speed of around $200\ \rm{km~s^{-1}}$ tentatively associated with loop retraction, as a guide to the outflow speed, but we only have a lower limit to $\theta$, as the dashed lines in Fig. \ref{fig:cartoon}e would represent the field in projection. { Using these values give $B \gtrsim 200$G}, more in line with expectation, { as photospheric magnetic fields up to 1000~G associated with the position of the ribbons are observed by SDO/HMI}. Using $L\sim 90$ arcseconds, the maximum { energy release is $\gtrsim 7.0\times 10^{11}A$ ergs}.  A nominal value of $A$, relative to the diameter of reconnecting loop strands of one arcsecond, would { give $\Delta E \sim 2.9 \times 10^{27}$~ergs per pair of strands}. From RHESSI imaging spectroscopic analysis, we found the total non-thermal energy of the electrons in the coronal source $E_\mathrm{non-thermal}=3.5 \times 10^{29}$ erg, integrating the impulsive phase (06:23:34 to 06:28:42~UT), which yields an energy rate of $\approx 10^{27}$ erg~s$^{-1}$. This would require the reconnection of { about one pair of strands every three seconds on average, or about 120 strands through the main impulsive phase}, corresponding to a `bundle' of strands around { 11\arcsec\ thick, about twice the estimated width of the coronal source}. The spreading motion seen in Figure \ref{fig:cartoon}f does still require an explanation; perhaps this is associated with the spreading of the reconnection site across a surface between two `sheets' of crossed magnetic field.  

\section{Conclusions}

We present clear observational evidence for the direct energy release and heating and possibly acceleration site during a non-eruptive two-ribbon flare. Combined RHESSI, Hinode/EIS, and SDO/AIA observational data reveal a coronal source, located roughly between the flaring ribbons, being heated impulsively. The coronal source shows fast downward plasma flows at 40--250 km/s at two different temperatures, measured through Doppler shifts of emission lines \ion{Fe}{xii} ($\log T = 6.2$) and \ion{Fe}{xxiv} ($\log T = 7.2$) by Hinode/EIS. RHESSI imaging spectroscopy reveals that the coronal source is suddenly heated, reaching a temperature peak of 12--13 MK about 10 seconds before the main HXR peak. Its fast cooling time is consistent with conductive losses, and implies that heating at the site reaches a peak before the peak of particle acceleration. Moreover, RHESSI HXR imaging shows that the coronal source dominates the footpoint emission at the peak of the impulsive phase. This is consistent with a scenario where plasma is first heated, and then electrons are accelerated in a turbulent and dense coronal source, with newly accelerated electrons are collisionally stopped before leaving the region. The energy deposited further heats the plasma, which then flows from the region at speeds around the sound speed towards each ribbon. These flows are observed as red-shifted emission lines by EIS and plasma motions by AIA images. This event does not show any eruptions and it is not associated with a CME. It may be explained by componend magnetic reconnection between current-carrying loops. In such a geometry, the energy that can be released seems to plausibly account for the flare energetics observed.

\begin{acknowledgements}
We would like to thank the anonymous reviewer for the comments and suggestions that helped to improve the paper. The research leading to these results has received funding from the European Community's Seventh Framework Programme (FP7/2007-2013) under grant agreement no. 606862 (F-CHROMA), from STFC grant ST/I001808/1 (PJAS, LF) and ST/L000741/1 (LF), and from an STFC `STEP' award to the University of Glasgow (DRG).
\end{acknowledgements}
\bibliographystyle{aa}
\bibliography{refs}

\begin{thebibliography}{58}
\expandafter\ifx\csname natexlab\endcsname\relax\def\natexlab#1{#1}\fi

\bibitem[{{Aschwanden} {et~al.}(1999){Aschwanden}, {Kosugi}, {Hanaoka},
  {Nishio}, \& {Melrose}}]{AschwandenKosugiHanaoka:1999}
{Aschwanden}, M.~J., {Kosugi}, T., {Hanaoka}, Y., {Nishio}, M., \& {Melrose},
  D.~B. 1999, \apj, 526, 1026

\bibitem[{{Aulanier} {et~al.}(2006){Aulanier}, {Pariat}, {D{\'e}moulin}, \&
  {DeVore}}]{AulanierPariatDemoulin:2006}
{Aulanier}, G., {Pariat}, E., {D{\'e}moulin}, P., \& {DeVore}, C.~R. 2006,
  \solphys, 238, 347

\bibitem[{{Brosius}(2012)}]{Brosius:2012}
{Brosius}, J.~W. 2012, \apj, 754, 54

\bibitem[{{Carmichael}(1964)}]{Carmichael:1964}
{Carmichael}, H. 1964, NASA Special Publication, 50, 451

\bibitem[{{Chae}(1999)}]{Chae:1999}
{Chae}, J. 1999, Journal of Korean Astronomical Society, 32, 137

\bibitem[{{Chifor} {et~al.}(2006){Chifor}, {Mason}, {Tripathi}, {Isobe}, \&
  {Asai}}]{ChiforMasonTripathi:2006}
{Chifor}, C., {Mason}, H.~E., {Tripathi}, D., {Isobe}, H., \& {Asai}, A. 2006,
  \aap, 458, 965

\bibitem[{{Culhane} {et~al.}(2007){Culhane}, {Harra}, {James}, {Al-Janabi},
  {Bradley}, {Chaudry}, {Rees}, {Tandy}, {Thomas}, {Whillock}, {Winter},
  {Doschek}, {Korendyke}, {Brown}, {Myers}, {Mariska}, {Seely}, {Lang}, {Kent},
  {Shaughnessy}, {Young}, {Simnett}, {Castelli}, {Mahmoud}, {Mapson-Menard},
  {Probyn}, {Thomas}, {Davila}, {Dere}, {Windt}, {Shea}, {Hagood}, {Moye},
  {Hara}, {Watanabe}, {Matsuzaki}, {Kosugi}, {Hansteen}, \&
  {Wikstol}}]{CulhaneHarraJames:2007}
{Culhane}, J.~L., {Harra}, L.~K., {James}, A.~M., {et~al.} 2007, \solphys, 243,
  19

\bibitem[{{Demoulin} {et~al.}(1993){Demoulin}, {van Driel-Gesztelyi},
  {Schmieder}, {Hemoux}, {Csepura}, \&
  {Hagyard}}]{Demoulinvan-Driel-GesztelyiSchmieder:1993}
{Demoulin}, P., {van Driel-Gesztelyi}, L., {Schmieder}, B., {et~al.} 1993,
  \aap, 271, 292

\bibitem[{{Dere} {et~al.}(1997){Dere}, {Landi}, {Mason}, {Monsignori Fossi}, \&
  {Young}}]{DereLandiMason:1997}
{Dere}, K.~P., {Landi}, E., {Mason}, H.~E., {Monsignori Fossi}, B.~C., \&
  {Young}, P.~R. 1997, \aaps, 125, 149

\bibitem[{{Dud{\'{\i}}k} {et~al.}(2014){Dud{\'{\i}}k}, {Janvier}, {Aulanier},
  {Del Zanna}, {Karlick{\'y}}, {Mason}, \&
  {Schmieder}}]{DudikJanvierAulanier:2014}
{Dud{\'{\i}}k}, J., {Janvier}, M., {Aulanier}, G., {et~al.} 2014, \apj, 784,
  144

\bibitem[{{Emslie}(1978)}]{Emslie:1978}
{Emslie}, A.~G. 1978, \apj, 224, 241

\bibitem[{{Graham} {et~al.}(2013){Graham}, {Hannah}, {Fletcher}, \&
  {Milligan}}]{GrahamHannahFletcher:2013}
{Graham}, D.~R., {Hannah}, I.~G., {Fletcher}, L., \& {Milligan}, R.~O. 2013,
  \apj, 767, 83

\bibitem[{{Grigis} \& {Benz}(2005)}]{GrigisBenz:2005}
{Grigis}, P.~C. \& {Benz}, A.~O. 2005, \apjl, 625, L143

\bibitem[{{Hara} {et~al.}(2006){Hara}, {Nishino}, {Ichimoto}, \&
  {Delaboudini{\`e}re}}]{HaraNishinoIchimoto:2006}
{Hara}, H., {Nishino}, Y., {Ichimoto}, K., \& {Delaboudini{\`e}re}, J.-P. 2006,
  \apj, 648, 712

\bibitem[{{Hara} {et~al.}(2011){Hara}, {Watanabe}, {Harra}, {Culhane}, \&
  {Young}}]{HaraWatanabeHarra:2011}
{Hara}, H., {Watanabe}, T., {Harra}, L.~K., {Culhane}, J.~L., \& {Young}, P.~R.
  2011, \apj, 741, 107

\bibitem[{{Hardy} {et~al.}(1998){Hardy}, {Melrose}, \&
  {Hudson}}]{HardyMelroseHudson:1998}
{Hardy}, S.~J., {Melrose}, D.~B., \& {Hudson}, H.~S. 1998, \pasa, 15, 318

\bibitem[{{Harra} {et~al.}(2005){Harra}, {D{\'e}moulin}, {Mandrini},
  {Matthews}, {van Driel-Gesztelyi}, {Culhane}, \&
  {Fletcher}}]{HarraDemoulinMandrini:2005}
{Harra}, L.~K., {D{\'e}moulin}, P., {Mandrini}, C.~H., {et~al.} 2005, \aap,
  438, 1099

\bibitem[{{Henoux} \& {Somov}(1987)}]{HenouxSomov:1987}
{Henoux}, J.~C. \& {Somov}, B.~V. 1987, \aap, 185, 306

\bibitem[{{Hirayama}(1974)}]{Hirayama:1974}
{Hirayama}, T. 1974, \solphys, 34, 323

\bibitem[{{Hudson}(1991)}]{Hudson:1991}
{Hudson}, H.~S. 1991, \solphys, 133, 357

\bibitem[{{Hurford} {et~al.}(2002){Hurford}, {Schmahl}, {Schwartz}, {Conway},
  {Aschwanden}, {Csillaghy}, {Dennis}, {Johns-Krull}, {Krucker}, {Lin},
  {McTiernan}, {Metcalf}, {Sato}, \& {Smith}}]{HurfordSchmahlSchwartz:2002}
{Hurford}, G.~J., {Schmahl}, E.~J., {Schwartz}, R.~A., {et~al.} 2002, \solphys,
  210, 61

\bibitem[{{Imada} {et~al.}(2013){Imada}, {Aoki}, {Hara}, {Watanabe}, {Harra},
  \& {Shimizu}}]{ImadaAokiHara:2013}
{Imada}, S., {Aoki}, K., {Hara}, H., {et~al.} 2013, \apjl, 776, L11

\bibitem[{{Joshi} {et~al.}(2007){Joshi}, {Manoharan}, {Veronig}, {Pant}, \&
  {Pandey}}]{JoshiManoharanVeronig:2007}
{Joshi}, B., {Manoharan}, P.~K., {Veronig}, A.~M., {Pant}, P., \& {Pandey}, K.
  2007, \solphys, 242, 143

\bibitem[{{Kontar} {et~al.}(2014){Kontar}, {Bian}, {Emslie}, \&
  {Vilmer}}]{KontarBianEmslie:2014}
{Kontar}, E.~P., {Bian}, N.~H., {Emslie}, A.~G., \& {Vilmer}, N. 2014, \apj,
  780, 176

\bibitem[{{Kopp} \& {Pneuman}(1976)}]{KoppPneuman:1976}
{Kopp}, R.~A. \& {Pneuman}, G.~W. 1976, \solphys, 50, 85

\bibitem[{{Landi} {et~al.}(2013){Landi}, {Young}, {Dere}, {Del Zanna}, \&
  {Mason}}]{LandiYoungDere:2013}
{Landi}, E., {Young}, P.~R., {Dere}, K.~P., {Del Zanna}, G., \& {Mason}, H.~E.
  2013, \apj, 763, 86

\bibitem[{{Lemen} {et~al.}(2012){Lemen}, {Title}, {Akin}, {Boerner}, {Chou},
  {Drake}, {Duncan}, {Edwards}, {Friedlaender}, {Heyman}, {Hurlburt}, {Katz},
  {Kushner}, {Levay}, {Lindgren}, {Mathur}, {McFeaters}, {Mitchell}, {Rehse},
  {Schrijver}, {Springer}, {Stern}, {Tarbell}, {Wuelser}, {Wolfson}, {Yanari},
  {Bookbinder}, {Cheimets}, {Caldwell}, {Deluca}, {Gates}, {Golub}, {Park},
  {Podgorski}, {Bush}, {Scherrer}, {Gummin}, {Smith}, {Auker}, {Jerram},
  {Pool}, {Soufli}, {Windt}, {Beardsley}, {Clapp}, {Lang}, \&
  {Waltham}}]{LemenTitleAkin:2012}
{Lemen}, J.~R., {Title}, A.~M., {Akin}, D.~J., {et~al.} 2012, \solphys, 275, 17

\bibitem[{{Lin} {et~al.}(2002){Lin}, {Dennis}, {Hurford}, {Smith}, {Zehnder},
  {Harvey}, {Curtis}, {Pankow}, {Turin}, {Bester}, {Csillaghy}, {Lewis},
  {Madden}, {van Beek}, {Appleby}, {Raudorf}, {McTiernan}, {Ramaty}, {Schmahl},
  {Schwartz}, {Krucker}, {Abiad}, {Quinn}, {Berg}, {Hashii}, {Sterling},
  {Jackson}, {Pratt}, {Campbell}, {Malone}, {Landis}, {Barrington-Leigh},
  {Slassi-Sennou}, {Cork}, {Clark}, {Amato}, {Orwig}, {Boyle}, {Banks},
  {Shirey}, {Tolbert}, {Zarro}, {Snow}, {Thomsen}, {Henneck}, {McHedlishvili},
  {Ming}, {Fivian}, {Jordan}, {Wanner}, {Crubb}, {Preble}, {Matranga}, {Benz},
  {Hudson}, {Canfield}, {Holman}, {Crannell}, {Kosugi}, {Emslie}, {Vilmer},
  {Brown}, {Johns-Krull}, {Aschwanden}, {Metcalf}, \&
  {Conway}}]{LinDennisHurford:2002}
{Lin}, R.~P., {Dennis}, B.~R., {Hurford}, G.~J., {et~al.} 2002, \solphys, 210,
  3

\bibitem[{{Litvinenko} \& {Somov}(1993)}]{LitvinenkoSomov:1993}
{Litvinenko}, Y.~E. \& {Somov}, B.~V. 1993, \solphys, 146, 127

\bibitem[{{Liu} {et~al.}(2012){Liu}, {Liu}, {T{\"o}r{\"o}k}, {Wang}, \&
  {Wang}}]{LiuLiuTorok:2012}
{Liu}, R., {Liu}, C., {T{\"o}r{\"o}k}, T., {Wang}, Y., \& {Wang}, H. 2012,
  \apj, 757, 150

\bibitem[{{Liu} \& {Fletcher}(2009)}]{LiuFletcher:2009}
{Liu}, S. \& {Fletcher}, L. 2009, \apjl, 701, L34

\bibitem[{{Liu} {et~al.}(2010){Liu}, {Han}, \&
  {Fletcher}}]{LiuHanFletcher:2010}
{Liu}, S., {Han}, F., \& {Fletcher}, L. 2010, \apj, 709, 58

\bibitem[{{Liu} {et~al.}(2013){Liu}, {Chen}, \&
  {Petrosian}}]{LiuChenPetrosian:2013}
{Liu}, W., {Chen}, Q., \& {Petrosian}, V. 2013, \apj, 767, 168

\bibitem[{{Longcope}(1996)}]{Longcope:1996}
{Longcope}, D.~W. 1996, \solphys, 169, 91

\bibitem[{{Melrose}(2004)}]{Melrose:2004}
{Melrose}, D. 2004, \solphys, 221, 121

\bibitem[{{Melrose}(1997)}]{Melrose:1997}
{Melrose}, D.~B. 1997, \apj, 486, 521

\bibitem[{{Milligan} \& {Dennis}(2009)}]{MilliganDennis:2009}
{Milligan}, R.~O. \& {Dennis}, B.~R. 2009, \apj, 699, 968

\bibitem[{{Narukage} {et~al.}(2014){Narukage}, {Shimojo}, \&
  {Sakao}}]{NarukageShimojoSakao:2014}
{Narukage}, N., {Shimojo}, M., \& {Sakao}, T. 2014, \apj, 787, 125

\bibitem[{{Parker}(1963)}]{Parker:1963}
{Parker}, E.~N. 1963, \apjs, 8, 177

\bibitem[{{Petschek}(1964)}]{Petschek:1964}
{Petschek}, H.~E. 1964, NASA Special Publication, 50, 425

\bibitem[{{Priest} \& {Forbes}(2000)}]{PriestForbes:2000}
{Priest}, E. \& {Forbes}, T. 2000, {Magnetic Reconnection}

\bibitem[{{Rosner} {et~al.}(1978){Rosner}, {Tucker}, \&
  {Vaiana}}]{RosnerTuckerVaiana:1978}
{Rosner}, R., {Tucker}, W.~H., \& {Vaiana}, G.~S. 1978, \apj, 220, 643

\bibitem[{{Scherrer} {et~al.}(2012){Scherrer}, {Schou}, {Bush}, {Kosovichev},
  {Bogart}, {Hoeksema}, {Liu}, {Duvall}, {Zhao}, {Title}, {Schrijver},
  {Tarbell}, \& {Tomczyk}}]{ScherrerSchouBush:2012}
{Scherrer}, P.~H., {Schou}, J., {Bush}, R.~I., {et~al.} 2012, \solphys, 275,
  207

\bibitem[{{Schwartz} {et~al.}(2002){Schwartz}, {Csillaghy}, {Tolbert},
  {Hurford}, {McTiernan}, \& {Zarro}}]{SchwartzCsillaghyTolbert:2002}
{Schwartz}, R.~A., {Csillaghy}, A., {Tolbert}, A.~K., {et~al.} 2002, \solphys,
  210, 165

\bibitem[{{Shen} {et~al.}(2014){Shen}, {Zhou}, {Ji}, {Wiegelmann}, {Inhester},
  \& {Feng}}]{ShenZhouJi:2014}
{Shen}, J., {Zhou}, T., {Ji}, H., {et~al.} 2014, \apj, 791, 83

\bibitem[{{Sim{\~o}es} {et~al.}(2013){Sim{\~o}es}, {Fletcher}, {Hudson}, \&
  {Russell}}]{SimoesFletcherHudson:2013}
{Sim{\~o}es}, P.~J.~A., {Fletcher}, L., {Hudson}, H.~S., \& {Russell}, A.~J.~B.
  2013, \apj, 777, 152

\bibitem[{{Sim{\~o}es} \& {Kontar}(2013)}]{SimoesKontar:2013}
{Sim{\~o}es}, P.~J.~A. \& {Kontar}, E.~P. 2013, \aap, 551, A135

\bibitem[{{Sturrock}(1966)}]{Sturrock:1966}
{Sturrock}, P.~A. 1966, \nat, 211, 695

\bibitem[{{Su} {et~al.}(2013){Su}, {Veronig}, {Holman}, {Dennis}, {Wang},
  {Temmer}, \& {Gan}}]{SuVeronigHolman:2013}
{Su}, Y., {Veronig}, A.~M., {Holman}, G.~D., {et~al.} 2013, Nature Physics, 9,
  489

\bibitem[{{Sweet}(1958)}]{Sweet:1958}
{Sweet}, P.~A. 1958, in IAU Symposium, Vol.~6, Electromagnetic Phenomena in
  Cosmical Physics, ed. B.~{Lehnert}, 123

\bibitem[{{Syrovatskii}(1966)}]{Syrovatskii:1966}
{Syrovatskii}, S.~I. 1966, \sovast, 10, 270

\bibitem[{{Tandberg-Hanssen} \& {Emslie}(2009)}]{Tandberg-HanssenEmslie:2009}
{Tandberg-Hanssen}, E. \& {Emslie}, A.~G. 2009, {The Physics of Solar Flares}

\bibitem[{{van Driel-Gesztelyi} {et~al.}(2014){van Driel-Gesztelyi}, {Baker},
  {T{\"o}r{\"o}k}, {Pariat}, {Green}, {Williams}, {Carlyle}, {Valori},
  {D{\'e}moulin}, {Kliem}, {Long}, {Matthews}, \&
  {Malherbe}}]{van-Driel-GesztelyiBakerTorok:2014}
{van Driel-Gesztelyi}, L., {Baker}, D., {T{\"o}r{\"o}k}, T., {et~al.} 2014,
  \apj, 788, 85

\bibitem[{{Wang} {et~al.}(2007){Wang}, {Sui}, \& {Qiu}}]{WangSuiQiu:2007}
{Wang}, T., {Sui}, L., \& {Qiu}, J. 2007, \apjl, 661, L207

\bibitem[{{Watanabe} {et~al.}(2010){Watanabe}, {Hara}, {Sterling}, \&
  {Harra}}]{WatanabeHaraSterling:2010}
{Watanabe}, T., {Hara}, H., {Sterling}, A.~C., \& {Harra}, L.~K. 2010, \apj,
  719, 213

\bibitem[{{Wheatland} \& {Melrose}(1995)}]{WheatlandMelrose:1995}
{Wheatland}, M.~S. \& {Melrose}, D.~B. 1995, \solphys, 158, 283

\bibitem[{{Yashiro} {et~al.}(2006){Yashiro}, {Akiyama}, {Gopalswamy}, \&
  {Howard}}]{YashiroAkiyamaGopalswamy:2006}
{Yashiro}, S., {Akiyama}, S., {Gopalswamy}, N., \& {Howard}, R.~A. 2006, \apjl,
  650, L143

\bibitem[{{Yokoyama} {et~al.}(2001){Yokoyama}, {Akita}, {Morimoto}, {Inoue}, \&
  {Newmark}}]{YokoyamaAkitaMorimoto:2001}
{Yokoyama}, T., {Akita}, K., {Morimoto}, T., {Inoue}, K., \& {Newmark}, J.
  2001, \apjl, 546, L69

\end{thebibliography}
\end{document}